\documentstyle[psfig]{basi}
\batchmode
\begin{document}

\title[GMRT observations of HII regions]{The electron temperatures of HII regions S~201, S~206, and S~209 : 
Multi-frequency GMRT observations}
\author[Omar, chengalur, \& Roshi]%
       {Amitesh Omar$^{1}$,  J.N. Chengalur$^{2}$, and D.A. Roshi$^{3}$ \\
        1. Raman Research Institute, Bangalore, India\\2. National Center for Radio Astrophysics (TIFR), Pune, India
\\3. National Radio Astronomical Observatory, Green Bank, USA }

\maketitle
\label{firstpage}

\begin{abstract} Three Galactic HII regions, viz., S~201, S~206, and S~209 have been imaged at three frequencies,
viz., 232, 327, and 610 MHz using the GMRT. The resolutions of these images are typically 15" at 232, 10" at 327,
and 6" at 610 MHz. These are the highest resolution low frequency images of these HII regions. We found that all
three HII regions have core--envelope morphologies. We use the high resolution afforded by the data to estimate the
electron temperatures of the compact cores of these HII regions. These estimates of the electron temperatures are
consistent with an increase in the temperature with Galacto-centric distance; an effect attributed to a decrease in
the heavy elements abundances at large Galacto-centric distances.  \end{abstract} 

\section{Introduction}

The radio recombination line measurements of HII regions in our Galaxy indicate a systematic increase in the
electron temperature with increasing Galacto-centric distance (Churchwell et al. 1978). This effect is believed to
be caused by a decrease in the heavy elements abundances with increasing Galacto-centric distance since metals are
primary cooling agents in HII regions. Consistent with the above hypothesis, various optical line studies of HII
regions, planetary nebulae, and supernova remnants have established a negative radial gradient of heavy elements in
the disk of the Milky-way (Henry \& Worthey 1999).  The variation in the electron temperature
(T$_{e}$) with the Galacto-centric distance (R$_{G}$) has been fitted using linear functions with slopes in the
range of 300--400 K kpc $^{-1}$.

\section{Results} The electron temperatures of three HII regions, viz., S~201, S~206, and S~209 have been estimated
using high resolution (pc--scale) radio continuum images at 232, 327, and 610 MHz using the GMRT. The results are
given in table~1. The estimated values of temperatures are in agreement with that predicted from a linear fit of
temperature vs. distance obtained by Pena et al. (2000). The 327 MHz images of S~209 are shown in figure~1. The
detailed analysis is presented elsewhere (Omar et al. 2002).

\begin{table}[h]
\begin{center}
\label{tab:HII}
\begin{tabular}{lcccc}
\hline
\hline
\bf{Name} & \bf{RA (1950)} & \bf{Dec (1950)} & \bf{R$_{G}$ (kpc)} & \bf{T$_{e}$ (K)} \\
\hline
S~201 & $02^{h}59^{m}20^{s}.1$ & $+60^{o}16'10"$ & 10.5 & $7070\pm1100$\\
S~206 & $03^{h}59^{m}24^{s}.0$ & $+51^{o}11'00"$ & 11.1 & $8350\pm1600$\\
S~209 & $04^{h}07^{m}20^{s}.1$ & $+51^{o}02'30"$ & 17.7 & $10855\pm3670$\\
\hline
\end{tabular} 
\end{center} 
\end{table}

\begin{figure}[h]
\begin{centering}
{\mbox {\psfig{file=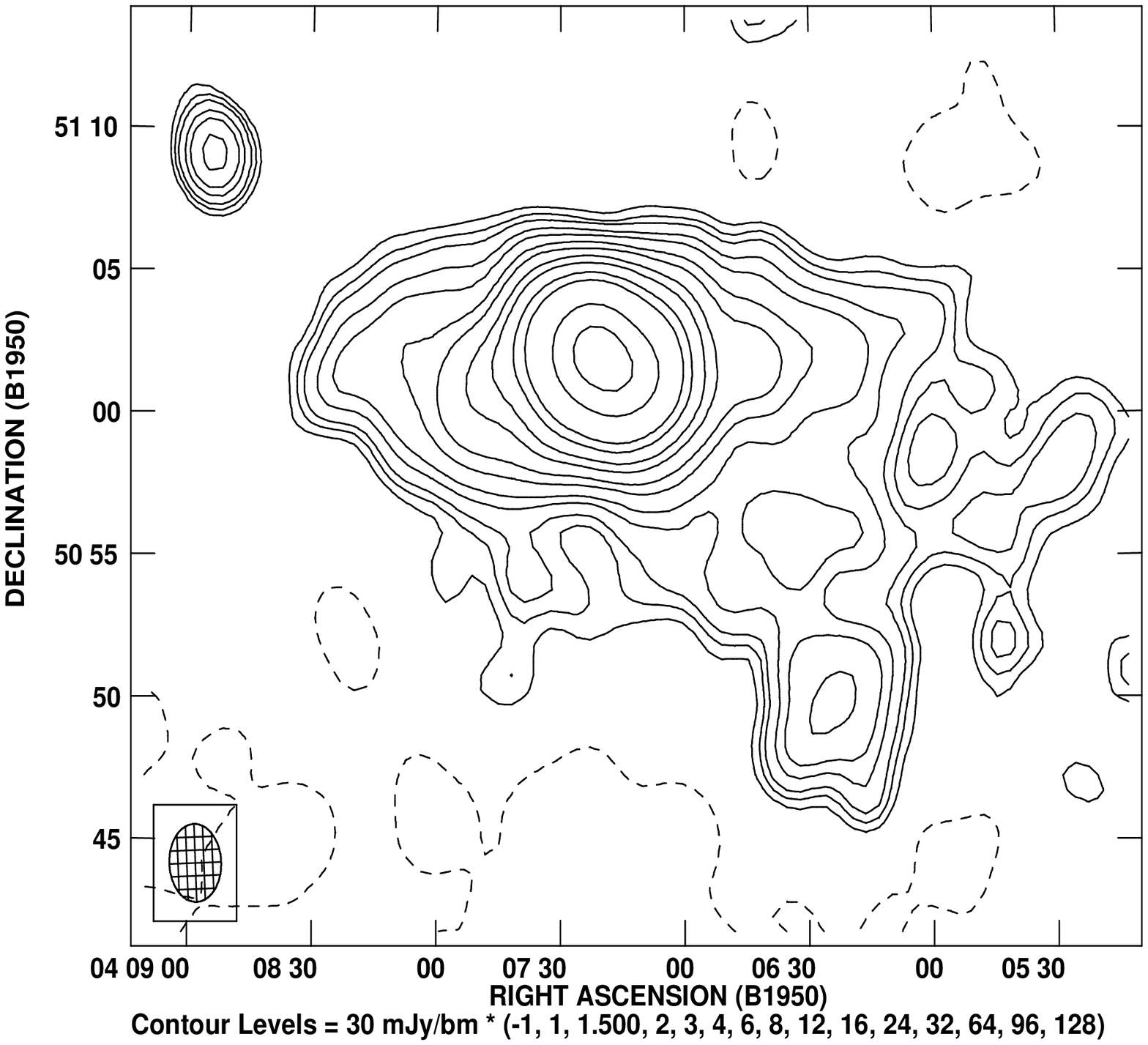,width=2.5truein,angle=0}}}
{\mbox {\psfig{file=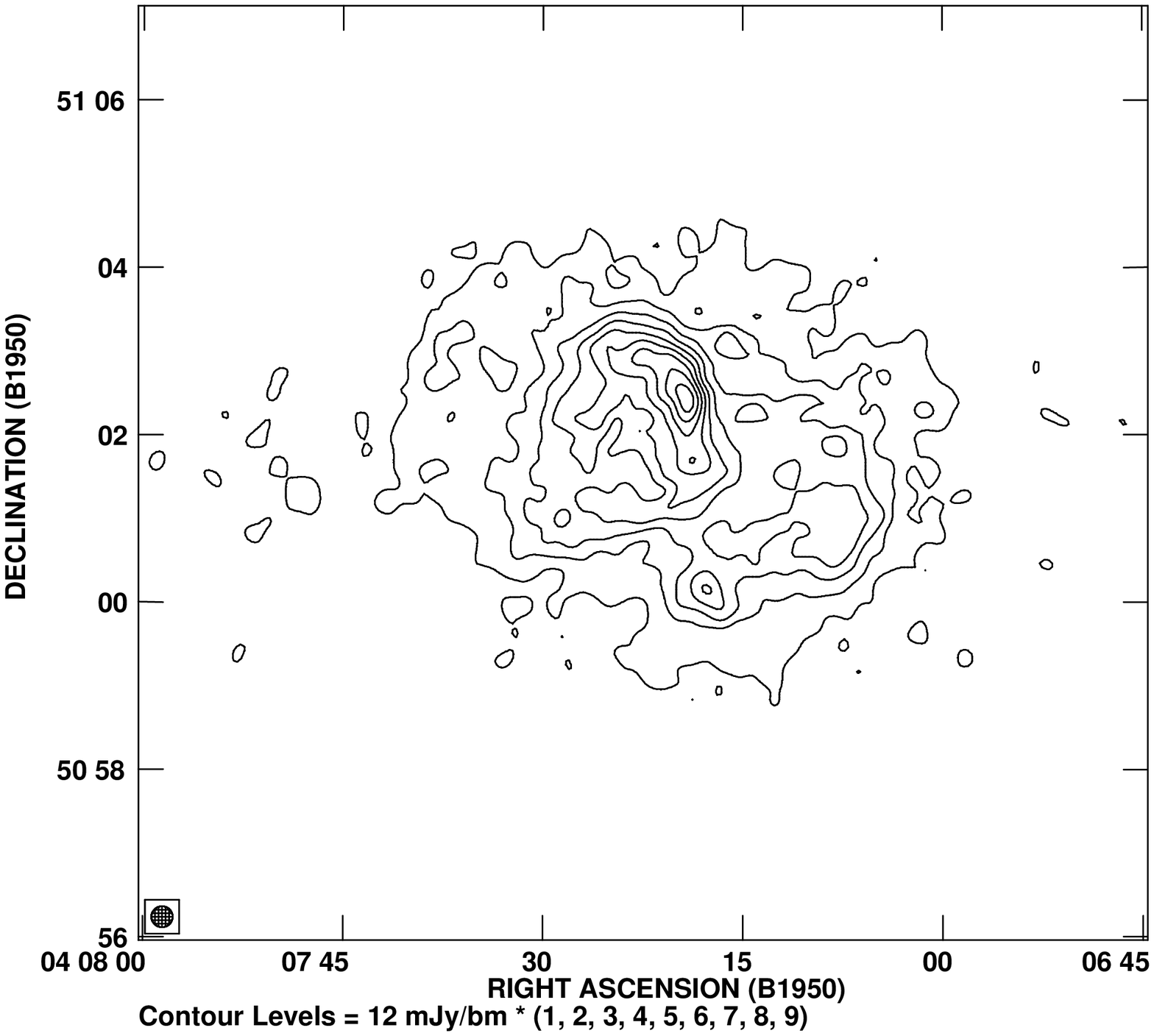,width=2.6truein,angle=0}}}
\caption{(left). 327 MHz GMRT image of S~209 made using only the central square antennas. The resolution is 
$164"\times119"$ and rms is 15 mJy beam$^{-1}$. (right). 327 MHz image of the core of S~209 with a 
resolution of $15"\times15"$ made using the full array. The rms in the image is 3 mJy beam$^{-1}$. }

\end{centering}
\end{figure}

\label{lastpage}

\end{document}